\documentclass{optica-article}
\journal{opticajournal}
\articletype{Research Article}
\usepackage{xr}
\usepackage{cleveref}
\usepackage{xcite}
\usepackage{lineno}
\usepackage[utf8]{inputenc}
\usepackage[T1]{fontenc}
\usepackage{graphicx}
\usepackage{siunitx} 
\usepackage{xcolor}
\usepackage{physics} 
\usepackage{empheq}
\usepackage{tabularx}
\usepackage{float}
\usepackage{svg}
\usepackage{hyperref}
\usepackage[separate-uncertainty=true]{siunitx}
\usepackage{soul}

\newcommand{\MC}[2]{\textcolor{black}{#2}}

\newcommand{\SD}[2]{\textcolor{black}{#2}}
\graphicspath{{Figures}}

\usepackage{soul}

\begin{document}
\title{Stabilizing nanoparticles in the intensity minimum: feedback levitation on an inverted potential.}
\author{Salamb\^{o} Dago\authormark{1}, J. Rieser\authormark{1}, M. A. Ciampini\authormark{1}, V. Mlynář\authormark{2}, A. Kugi\authormark{2,3}, M. Aspelmeyer\authormark{1,4}, A. Deutschmann-Olek\authormark{2}, N. Kiesel\authormark{1}}

\address{\authormark{1} University of Vienna, Faculty of Physics, Vienna Center for Quantum Science and Technology (VCQ), Vienna, Austria\\ \authormark{2} Automation and Control Institute (ACIN), TU Wien, Vienna, Austria\\\authormark{3} AIT Austrian Institute of Technology, Vienna, Austria\\
\authormark{4} Institute for Quantum Optics and Quantum Information (IQOQI) Vienna, Austrian Academy of Sciences, Vienna, Austria}

\email{\authormark{*}salambo.dago@univie.ac.at}

\begin{abstract*}
We demonstrate the stable trapping of a levitated nanoparticle \SD{on top}{at the apex} of an inverted potential using a combination of optical readout and electrostatic control. The feedback levitation on an inverted potential (FLIP) method stabilizes the particle at an intensity minimum.
By using a Kalman-filter-based linear-quadratic-Gaussian (LQG) control method, we confine a particle to within $\sigma_x=\SI{9\pm0.5}{nm}$ of the potential maximum at an effective temperature of \SI{16\pm1}{K} in a room-temperature environment. Despite drifts in the absolute position of the potential maximum, we can keep the nanoparticle at the apex by estimating the drift from the particle dynamics using the Kalman filter. Our approach may enable new levitation-based sensing schemes with enhanced bandwidth. It also paves the way for optical levitation at zero intensity of an optical potential, which alleviates decoherence effects due to material-dependent absorption and is hence relevant for macroscopic quantum experiments. 
\end{abstract*}
\section{Introduction}

Optical levitation of mesoscopic dielectrics has emerged as a vibrant and promising field, with the potential to advance sensing technologies~\cite{Goldwater2019,Millen2015}, explore fundamental aspects of stochastic and quantum thermodynamics \cite{Gieseler2018}, and probe macroscopic quantum physics~\cite{Millen2020,gonzalez-ballestero_levitodynamics_2021,Romero-Isart_2017}. The approach offers valuable features, including the flexible spatio-temporal manipulation of the potential landscape~\cite{Ciampini2021,Neumeier2024}, exceptional isolation from the environment \cite{gonzalez-ballestero_levitodynamics_2021}, and the availability of optomechanical interactions \cite{Aspelmeyer2014}. These features enable the creation of pure quantum states through ground-state cooling using optical resonators~\cite{Delic2020,Ranfagni2022,Piotrowski2023} or near-Heisenberg-limited continuous position monitoring in free space combined with feedback control \cite{Tebbenjohanns2021,Magrini2021,kamba_nanoscale_2023}.

Quantum feedback control has been successfully implemented on several experimental platforms, for example, \cite{Sayrin2011,Magrini2021,Melo2024}. In the context of levitating solids, it promises to be a powerful tool for manipulating quantum states of motion beyond mere ground-state cooling \cite{Ralph2018}. While conventional levitation schemes rely on the confining optical forces coinciding with intensity maxima of light fields, electronic feedback control based on optical readout can stabilize and cool particle motion regardless of the position within the optical potential. For instance, a levitated object can be suspended within the dark region of an optical mode, characterized by a zero-crossing of the electric field. This configuration enables optical readout with minimized absorption. Stabilization can replace relying on low-field seeking particles \cite{Melo2020}. Once implemented, such an approach may allow the levitation of a broader range of materials, such as metal particles or diamond-nanoparticles with NV-centers that are heating in conventional optical tweezers to the level of destruction even at relatively low vacuum levels \cite{Jin2024,Delord2020,Conangla2018,Frangeskou2018,riviere_thermometry_2022}. In addition, minimizing the heating due to absorption is crucial for advancing to larger quantum superposition states, even with transparent dielectrics, to minimize decoherence from black-body radiation. Consequently, this method may complement conventional dark electrostatic traps \cite{Yin2013,Monteiro2020,Moore2021,Dania2024}. However, the feasibility of active feedback stabilization in optical levitation schemes away from the potential minimum has not yet been demonstrated.
Here, we take a step in this direction by demonstrating the stabilization of a levitated nanoparticle at the \SD{top}{apex} of the inverted potential. This scenario is analogous to the classic inverted pendulum problem in dynamics and control theory. Using a Kalman filter in combination with non-adaptive feedback approach, we achieve stable confinement and cooling of the nanoparticle motion. Levitation at the apex of the potential is particularly sensitive to experimental drifts, such as beam-pointing fluctuations, which affect the position of the optical potential with respect to the detection (used by the feedback). To mitigate this issue, we implement an adaptive feedback approach based on a simple extension of the standard Kalman filter that autonomously compensates for such drifts.

\section{Experimental platform}
We work with a sub-micrometer silica particle (diameter $d = \SI{210}{nm}$, mass $m \sim \SI{1e-17}{kg}$) in a vacuum chamber at room temperature ($T_0 = \SI{300}{K}$) and pressure of $P = \SI{1.5}{mbar}$, corresponding to a friction coefficient $\gamma \simeq 2\pi \times \SI{0.8}{kHz}$. We create an optical double-well potential transverse (x-direction) to the joint optical axis formed by overlapping a TEM00 mode and a TEM10 mode of the incoming laser beam\cite{Ciampini2021} (refer to Supp. Mat. section 1 for details). The beam axes are collinear. The beam waists along the x-axis are approximately $\sim \SI{730}{nm}$ \SD{}{ and $\sim \SI{660}{nm}$\, for polarization along and orthogonal to the axis respectively. This is determined via the respective transversal particle frequencies\cite{Millen2020}(see Supp. Mat. section 1 for details) }. The two optical modes are orthogonally polarized and frequency-detuned to avoid interference (see Fig.~ S1 Sup. Mat). Along the beam axis (z-direction) and in the other transverse (y) direction, the trap is harmonic. Let us briefly note that the \SD{top (or apex)}{apex} of the 1D inverted potential (centered around $x=0$) corresponds to a local intensity minimum and could alternatively be designated as the center of the double well potential or as the unstable local potential maximum. We use optical powers of $\SI{320}{mW}$ and $\SI{540}{mW}$ for the TEM00 and TEM10 modes, respectively. This configuration results in optical frequencies of $f_{x,\mathrm{IP}} = \SI{-50}{kHz}$, $f_y = \SI{160}{kHz}$, and $f_z = \SI{48}{kHz}$ at the unstable potential maximum at the center of the double well, and $f_{x,\pm} = \SI{65}{kHz}$ at the stable trapping positions of the two wells, located at a distance of approximately $\pm\SI{200}{nm}$ from the beam axis.  Note that we define the negative resonance frequencies via a negative stiffness following the general expression: \SD{}{$f = \textrm{sgn}(k) \sqrt{|k|/m} /(2\pi)$}, for any harmonic potential $U(x) = \frac{1}{2} k x^2$ (attractive or inverted). The optical parameters of the inverted potential, including stiffness and harmonic oscillator resonance frequencies, are summarized in Table~\ref{table_param}.

\begin{figure}[H]
	\centering
	\includegraphics[width=11cm]{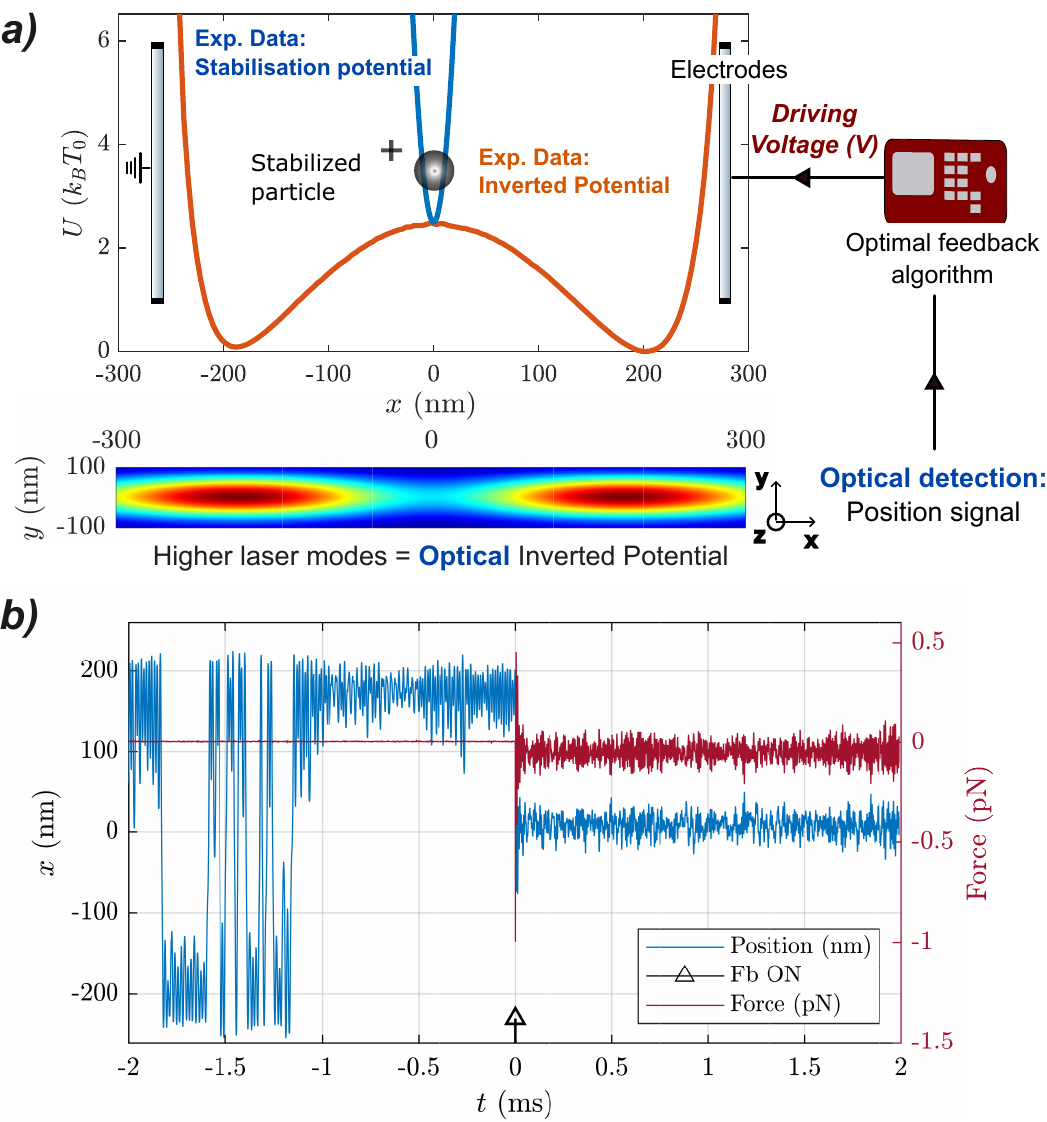}
	\caption{\textbf{(a) Schematic representation of the experiment}: We create an optical double-well potential (orange curve) transversal to the common beam axis of a TEM00 and a TEM10 mode, which have different frequencies and orthogonal polarizations. A charged nanoparticle \SD{}{(charge symbolised by a $+$ sign)} is stabilized at the central potential maximum (intensity minimum; $x=x_{\mathrm{apex}}=0$) using a feedback force applied via a driving voltage to a pair of electrodes. Based on the optical readout of the TEM00 mode in forward detection, the feedback control is processed with a linear quadratic regulator. The feedback force localizes the particle motion in an effective harmonic trap (blue curve) near the \SD{top}{apex} ($x_{\mathrm{apex}}=0$) of the inverted potential locally harmonic with a negative spring constant (orange curve). \SD{}{The effective stabilization potential (blue) and the inverted potential, (orange) are computed from the corresponding position probability distribution ($\mathrm{PDF}(x)$ assuming a Gibbs equilibrium distribution : $U_\mathrm{eff}(x)=U_0+k_BT_0\ln{\mathrm{PDF}(x)}$. The probability distributions are determined from a $\SI{2.5}{ms}$ steady state time trace with and without feedback, respectively. For clarity the potential energy offset is chosen to match the blue and orange curves at $x=0$.}
\textbf{(b) Feedback force and position trajectories:} \MC{The plot shows the p}Particle position  ($x$, blue) and applied electrostatic force (red) over time. Without feedback control ($t<0$, Force: $\pm\SI{0}{pN}$), the particle's free dynamics exhibit jumps between the two wells, corresponding to a bimodal probability distribution (see Fig.~\ref{PS}). When feedback control is activated ($t>0$), the particle ceases jumping, and its trajectory shows damped harmonic motion near the \SD{tip}{apex} of the centered inverted potential (see Fig.~\ref{PS}).}
	\label{Time Trace}
\end{figure}

To detect nanoparticle motion, we first isolate light at the polarization of the TEM00 mode. This enables position readout of the x and z motion in a standard detection scheme based on interference between the TEM00 mode and the light scattered by the particle \cite{Gieseler2012}. To prevent drifts in position detection along the x-axis, the TEM00 beam is actively stabilized at the entrance of the chamber using piezo-controlled mirrors. This stabilization also defines the null-position of the particle along the x-axis as the center of the TEM00 mode. The detection signal is processed via an FPGA and used to apply feedback control to the charged silica nanosphere electronically, via a pair of electrodes mounted along the x-axis, as illustrated in Fig.~\ref{Time Trace}a.

\begin{table}
  \centering
\begin{tabular}{|p{0.17\columnwidth}|p{0.2\columnwidth}|p{0.19\columnwidth}|p{0.17\columnwidth}|p{0.17\columnwidth}|}
\hline  \centering  Axis of motion
& \centering x $\approx$~ $\pm\SI{200}{nm}$ & \centering x $\approx$~$\SI{0}{nm}$ &\centering  $y$ (x $\approx$~$\SI{0}{nm}$) & \centering\arraybackslash $z$ (x $\approx$ $\SI{0}{nm}$) \\
\hline
$f$ (kHz)&$f_{x,\pm}=\SI{65}{}\pm1$&$f_{x,\mathrm{IP}}=\SI{-50}{}\pm 3$&$f_y=\SI{160}{}\pm1$&$f_z=\SI{48}{}\pm1$ \\
\hline
$k$($\mu$N/m) &$k_{x,\pm}=\SI{1.60}{}\pm0.05$&$k_{x,\mathrm{IP}}=\SI{-1.0}{}\pm0.1$&$k_y=\SI{10.0}{}\pm0.1$&$k_z=\SI{0.90}{}\pm0.04$ \\
\hline
\end{tabular}

\caption{
\textbf{1D Optical Inverted Potential Parameters} for the experimental values used in this work (main text). Along the \(x\)-axis, the stiffness at the center (\(k_{x,\mathrm{IP}}\)) and near the potential minima in the two wells (\(k_{x,\pm}\)) are calibrated through harmonic fits of the potential at each position (error bars are inferred from the fit, and propagated to $f_{x,\mathrm{IP}}$), which are inferred from the particle probability density function (PDF) without feedback control. The corresponding frequencies in the wells (\(f_{x,\pm}\)) are independently determined from the motional spectra (from the Lorentzian fit of the resonance peaks). The frequencies and spring constants along the other axes are similarly obtained from their respective resonance frequencies at the center of the potential (see Supplementary Material, Fig.~ S2).
}

 \label{table_param}
\end{table}

\section{Feedback control}

Balancing by feedback on an unstable point of the nano-oscillator recalls the dynamic stabilization of an inverted pendulum, which traces back to the 1960s \cite{LUNDBERG2010131}. Here, the feedback acts strictly as an external (electrostatic) force, contrary to the so-called \textit{parametric} feedback, which acts on the laser beam intensity, i.e. the stiffness of the potential.

The optimal feedback control scheme combines a Kalman filter that estimates the state of the system with a Linear Quadratic Regulator (LQR) that generates the feedback signal. This Linear Quadratic Gaussian (LQG) algorithm has been used for optimal cooling of the motion of a levitated nanoparticle in a confining harmonic potential \cite{Magrini2021}. While it assumes linear dynamics, it does not require confinement (in the form of an attractive potential). Thus, it can be applied equally to repulsive harmonic dynamics with a negative spring constant investigated here. 
Neglecting the other directions of motion, our system can be modelled by the 1D evolution of a particle of mass $m$ evolving in a potential $U(x)$, using the Langevin Equation 
\begin{equation}
m\ddot{x}+\gamma \dot{x}+\frac{dU}{dx}=F_{\mathrm{th}}+u
\end{equation}
 where  $\gamma$ is the friction coefficient of the environment, \textit{u} is the feedback force applied through the electrodes, and $F_{\mathrm{th}}$ the thermal noise satisfying $\langle F_{\mathrm{th}}(t) F_{\mathrm{th}}( t^\prime) \rangle=2 \gamma k_B T_0 \delta(t-t^\prime)$ at room temperature $T_0$, with the Boltzmann constant $k_B$. Finally, we express the potential as $U(x)=\frac{1}{2}k_{x,\mathrm{IP}}(x-x_{\mathrm{apex}})^2$.
 
 We approximate the double well potential near its apex by a repulsive quadratic potential whose inverted stiffness $k_{x,\mathrm{IP}}$ is determined by the power ratio of the two respective beams. We also incorporate $x_{\mathrm{apex}}$ to describe the distance of the apex of the potential to the zero-point of our detection, to account for its slow drifts caused by beam-pointing.   
In the LQG approach, the feedback force $u$ is chosen as a function of the state-variable vector $\xi$, such as $u=\mathbf{K}\hat{\xi}$, where $\hat{\xi}$ represents the Kalman filter estimates of the state variables, e.g. particle position and velocity. For an infinite time horizon, the control parameters (or feedback gain) $\mathbf{K}$ of the optimal control law are obtained by solving the Algebraic Riccati Equation (ARE), see also \cite{Magrini2021}. For this system, the solution gives such feedback parameters to minimize the velocity and displacement of the particle while balancing the feedback effort.

In our experimental conditions, the assumptions made on the potential shape (inverted harmonic, enforcing Gaussian distributions) as the basis of the LQG hold as long as the particle remains in the $\pm\SI{100}{nm}$ region around the apex. Within those bounds, the non-linear component of the repulsive potential is sufficiently small compared to the linear component. Furthermore, the detection sensitivity is constant when the particle remains in this range, close enough to the center of the detection beam.
\SD{Exiting this region will, as any mismatch between the feedback model and the actual dynamics, reduce the efficiency of the LQG control and limit the stabilization performance.}{If the particle escapes this region, the actual dynamics of the system will not be represented effectively by the feedback model, thus reducing the efficiency of the LQG control and limiting the stabilization performance.}

In the straightforward application of the LQG algorithm to stabilize the nanoparticle \SD{on top}{at the apex} of an inverted harmonic potential, a Kalman filter estimates the position and velocity of the particle in one axis, and these estimated states are used in the state feedback law for $u$. All other parameters of the model are assumed to be constant and known, and the remaining degrees of freedom are perfectly decoupled. Henceforth, this control concept will be called the \textbf{non-adaptive}  approach. 

In reality, the assumption of constant and known parameters does not hold, particularly for $x_{\mathrm{apex}}$, which drifts, thus leading to an additional term driving the particle. In order to address this issue, $x_{\mathrm{apex}}$ is also estimated in addition to the 1D particle position and velocity. Since the drift of $x_{\mathrm{apex}}$ occurs only on a time scale slower than 50 ms [30], the Kalman filter update for $x_{\mathrm{apex}}$ can be performed more slowly compared to the state variables $x$ and $\dot x$. Moreover, the estimation quality of $x_{\mathrm{apex}}$ is reduced by crosstalk in the detection signal between the z-motion at lower frequency and the x-motion (see details of Supp. Mat. Section 6.3). To account for this effect, we implement an independent detection of the z-motion in the experiment and add the position and velocity in the z-direction to the state vector. Accordingly, the state vector contains five components estimated by the Kalman filter: the four phase space variables in the x-z plane and the \SD{tip}{apex} position $(x_{\mathrm{apex}})$. The LQR operates to generate the feedback signal along the double-well potential, but is provided with the improved estimate of the particle motion in the x-direction relative to the potential \SD{tip}{apex} (\( x_e=x-x_{\mathrm{apex}} \) and \( \dot{x}_e \)).

\subsection{Stabilization performance indicators}

In a perfect scenario, without any drifts of $x_{\mathrm{apex}}$, the success of the stabilization could be evaluated by characterizing the confinement of the particle motion around $x=0$. However, as discussed above, the reality of our experiment includes such drifts. We characterize the quality of the stabilization \textbf{at the local potential maximum} using additional success indicators (for further discussion, see also Supp. Mat. Section 6:

\begin{enumerate}

     \item \textbf{Unimodal position probability distribution function (PDF):} the PDF of the particle position exhibits a single maximum. In particular, it does not show maxima caused by the two wells of the potential located at $\pm\SI{200}{nm}$ (Fig.~\ref{Comp_Feedbacks}a and c).
     
    \item \textbf{\SD{Flat $x$ spectrum}{Absence of x resonance peak}:} The power spectral density (PSD) of position does not exhibit the resonance frequency of the potential minima on either side of the double well (approximately $\sim \SI{65}{kHz}$). The power spectral density is shown in Supp. Mat. Fig.~S3.
  
    \item \textbf{Zero-mean driving force:} the particle is truly stabilized at the minimum intensity. If the particle were at a slope of the potential, the feedback would continuously compensate for a DC force corresponding to the local gradient of the potential, in addition to the stochastic forces acting on the particle. This criterion is evaluated at $\SI{2}{ms}$ time intervals, which corresponds to three times the feedback control time constant of $\SI{0.6}{ms}$ for adaptation to changes in the state vector (assuming the slowest scenario of the \textbf{adaptive approach}). The corresponding analysis is shown in Fig.~\ref{Comp_Feedbacks}b and d. 
   
\end{enumerate}

\section{Results}

We demonstrate two stabilization scenarios for comparison. First, the simple application of the LQG (combining Kalman filter and LQR), similar to what has been used in harmonic potentials (\textbf{non-adaptive} approach) \cite{Magrini2021}; second, an extended version (\textbf{adaptive} approach), where the controller follows the drift of the position of the potential maximum $x_{\mathrm{apex}}(t)$ relative to the X~detection readout.

\begin{figure}[H]
	\centering
	\includegraphics[width=\columnwidth]{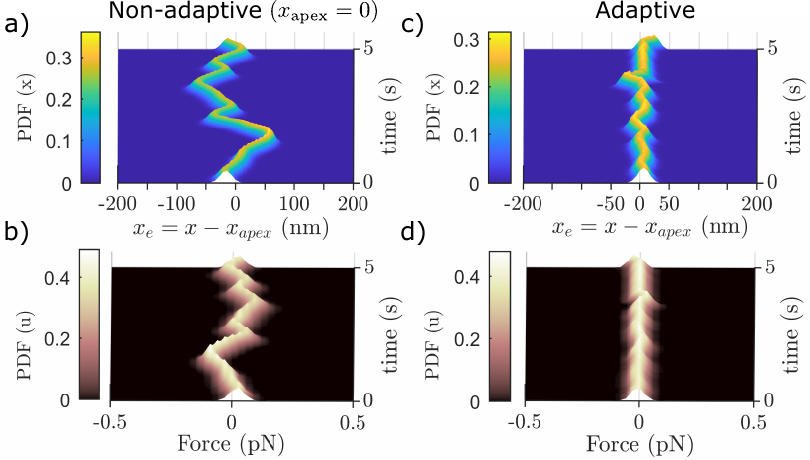}
\caption{\textbf{Temporal Evolution of the Probability Density Functions for Position and Control Force over 5 seconds}. The probability density functions (PDFs) for position (PDF($x$)) and control force (PDF($u$)) are obtained via normalized histograms based on 16 ms time windows in respectively the position and force signals. Note that PDF($x$) is displayed in the \SD{tip}{apex} frame of reference $x_e=x-x_{\mathrm{apex}}$. \MC{}{The white distribution indicates the initial PDF.} (a) The position PDF is unimodal, indicating confinement around $x=0$ (coinciding in the Static case with \SD{}{the assumption} $x_{\mathrm{apex}} = 0$ \SD{}{for lack of a drift estimate}).  (b) The control force PDF shows a slow drift, revealing the limit of the Static LQG efficiency (c) An adaptive algorithm estimates $x_{\mathrm{apex}}(t)$, and stabilizes the particle, resulting in a unimodal position distribution around the apex of the inverted potential (at $x_e=0$ in the \SD{tip}{apex} frame of reference). Details on the apex drift in the detection frame of reference are shown in Supplementary Material, \MC{}{(}Fig.~S4a and b\MC{}{)}. (d) The PDF in driving force is now centered near zero with minor drift.}
	
	\label{Comp_Feedbacks} 
\end{figure}

The results obtained by applying the simple \textbf{non-adaptive} approach are summarized in Fig.~\ref{Comp_Feedbacks}a and b, showing the PDFs in position and force during a $\SI{5}{s}$ stabilization time trace. The protocol allows confinement of the particle around $x=0$ (coinciding with $x_{\mathrm{apex}}=0$ in the model of \textbf{non-adaptive} approach), away from potential minima (located at $\pm \SI{200}{nm}$), getting a unimodal distribution (criterion 1). Also, the spectrum does not exhibit any dynamics beyond the expected effective harmonic motion (criterion 2), i.e., the \textbf{non-adaptive} approach scheme successfully flattens the resonance peak in the position Power Spectral Density (PSD), as shown in Supp. Mat. Fig.~S3. Nevertheless, the mean of the PDF in force drifts significantly in time, demonstrating that the particle is stabilized despite a drifting potential maximum - the particle is not at the local intensity minimum at all times (criterion 3 not fulfilled).

\begin{figure}
	\centering
	\includegraphics[width=11cm]{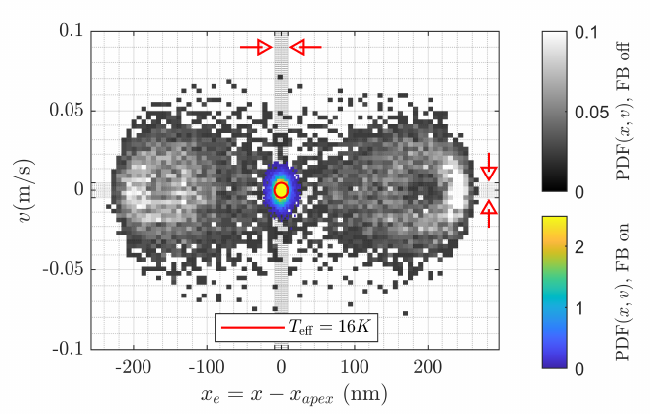}
\caption{\textbf{Normalized Phase Space Distributions: Free vs. Stabilized Dynamics}. The phase space probability density function (PDF($x, v$)) of a stabilized trajectory (shown in color) is overlaid on the bi-modal free dynamics (shown in gray). Both are derived from $\SI{2.5}{ms}$ time traces and displayed in the \SD{tip}{apex} position frame of reference, $x_e=x-x_{\mathrm{apex}}$ (with $x_{\mathrm{apex}}$ estimated by the feedback for the stabilized trajectory). The red circle and arrows indicate the standard deviations in position ($\sigma_x$) and velocity ($\sigma_v$) during the stabilized dynamics. Initially, the particle's motion spans $\sigma_x^{\textrm{free}} = \SI{100}{nm}$ near each potential well. The \textbf{adaptive} control reduces this spread to $\sigma_x = \SI{9}{nm}$ around the position of the \SD{tip}{apex}. Additionally, the control algorithm reduces the motion's effective temperature from $T_0 = 300 \text{ K}$ to $T_{\textrm{eff}} = m\sigma_v^2/k_B = \SI{16}{K}$, as indicated by the narrow distribution around the velocity axis.}
	\label{PS}
\end{figure}

The results of the \textbf{adaptive} approach are shown in Figs.~\ref{Comp_Feedbacks}c and d. The PDF in position demonstrates a narrow unimodal distribution complying with criterion 1 and exhibits no spectral features of the wells (criterion 2, see Supp. Mat. Fig.~S3). Moreover, in contrast to the simpler model, the particle is strongly confined near the \SD{tip}{apex} position (at $x_e=0$ in the \SD{tip}{apex} position frame of reference). \textcolor{black}{The main improvement is that the particle is continuously stabilized at the drifting apex of the potential. This is certified by the observation that the \SD{mean PDF in force}{of the force's probability distribution} is steady around 0 (criterion 3)}. With the adaptive approach, the motion of the particles is confined to $\sigma_x = \SI{9\pm0.5}{nm}$ around the position \SD{tip}{apex} at an effective temperature $T_{\textrm{eff}} = m\sigma_v^2/k_B = \SI{16\pm1}{K}$, where $\sigma_v^2$ is the variance of the velocity of the particles. The variances of the particle's position and velocity are computed directly from $\SI{5}{s}$ time-trace samples, with the error estimate deduced from repeating the stabilization procedure several times. Fig.~\ref{PS} shows the corresponding phase space probability distribution compared to the thermal distribution in a double well at \SI{300}{K}. This data is computed from a $\SI{2.5}{ms}$ detection sample. The experimentally determined bimodal PDF(x,v) of the free dynamics is shown as a black-and-white color map.

\section{Conclusion}

We successfully stabilize a nanoparticle at the \SD{top}{apex} of an optical inverted potential (intensity minimum)  using a Kalman filter and a linear quadratic regulator. Unlike previous work in harmonic potentials, the dynamic instability must be counteracted by the feedback. An extended feedback scheme also estimates and compensates for the experimental imperfections. We demonstrated that the experimental drifts of the potential maximum can be tracked, allowing the particle to follow the apex, transferring another control engineering method to the field of optomechanics with levitating particles. Quantum dynamics generated by inverted potentials are also of great interest to speed up the coherent dynamics needed to prepare and probe quantum macroscopic superpositions\cite{Romero-Isart_2017}.

Besides, the control scheme at a point of instability paves the way for particle stabilization at zero optical intensity, offering a route to the levitation of absorbing materials at high vacuum. Achieving this requires extending the stabilization method to 3D motion, as an intensity minimum is unstable in all three directions. The optical readout at these positions is feasible but requires novel detection schemes (Ref.~\cite{ORI2010}) and electrostatic control in all three directions.

Once extended to 3D, the feedback levitation on an inverted potential (FLIP) approach offers a path to quantum state control with optical readout yet minimized absorption, complementing dark trap schemes based on electrostatic or magnetic traps. This may enable the preparation of squeezed quantum states of nanoparticles with a cold bulk temperature, reducing black body decoherence, which currently limits schemes for matter-wave interference \cite{Neumeier2024,PhysRevLett.132.023601,bateman_near-field_2014}.

\begin{backmatter}

\bmsection{Acknowledgements}
We thank Oriol-Romero-Isart and Patrick Mauerer for enlightening discussions on the FLIP project and Gregor Thalhammer for support on the optical setup. This research was funded in whole or in part by the Austrian Science Fund (FWF) [10.55776/COE1] and [ 1055776/P36236]. For open access purposes, the author has applied a CC BY public copyright license to any author-accepted manuscript version arising from this submission. This project has received funding from the European Research Council (ERC) under the European Union’s Horizon 2020 research and innovation programme (grant agreement No 951234, Q-Xtreme) and by the European Union (HORIZON TMA MSCA Postdoctoral Fellowships - European Fellowships, FLIP, No 101106514). Views and opinions expressed are however those of the author(s) only and do not necessarily reflect those of the European Union or the European Commission-EU. Neither the European Union nor the granting authority can be held responsible for them. We acknowledge support from the Erwin Schrödinger Center for Quantum Science and Technology (ESQ) via a Discovery Grant.

\bmsection{Disclosures}
The authors declare no conflict of interest.

\bmsection{Supplemental document}
See Supplement 1 for supporting content. 
\end{backmatter}

\bibliography{Inverted}

\end{document}


\section{Experimental set-up}
\label{SuppExpSetUp}
\begin{figure}[!ht]
\includegraphics[width=12cm]{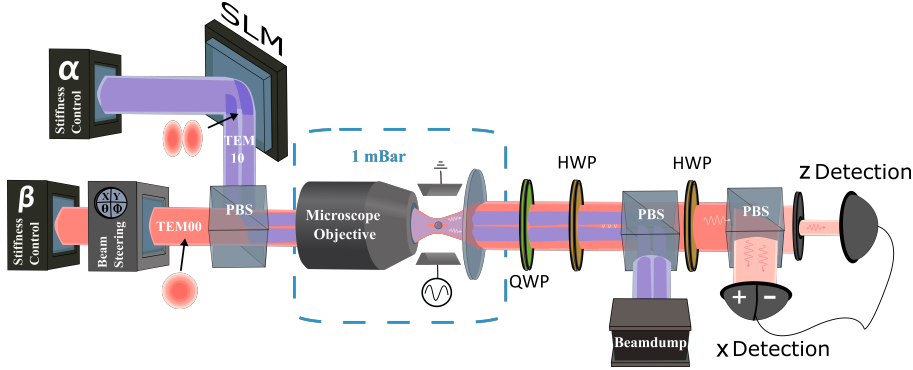}
\caption{ \textbf{Sup. Mat.: Setup used to generate the optical double well}. Two 1064nm laser beams are combined using a polarizing beamsplitter (PBS) and focussed using a 0.8 NA microscope objective. In the purple beam path, a spatial light modulator (SLM Meadowlark Inc.) applies a phase mask that converts the beam to a TEM10 mode in the Fourier plane. The intensity of the beams is controlled using acousto-optic modulators, allowing the potential to be shaped as desired. To reduce drifts, the red beam path is pointing stabilized. Radial electrodes allow the application of electrostatic forces on trapped particles.
A lens collects the light following the trap, which is used for detection. The TEM10 mode is filtered at a PBS, a half-wave and quarter-wave plate to control polarization. A combination of radial split mirror self-homodyne detection, and axial detection using a pinhole for mode selection is employed.} 
\label{SLMPot}
\end{figure}

 Two 1064nm laser beams generate a double well potential in the trapping area of a 0.8 NA microscope objective. \SD{}{To determine the waist of the two focused beams, we operate both beams as TEM00 modes, i.e., we set the SLM phase pattern to be flat. In both TEM00 modes the waist (consistent with the generic $1/e^2$ definition)  is calculated from the radial frequency and the optical power measured at the focus\cite{Millen2020}.}
 To avoid interference, the beams are orthogonally polarized and detuned in frequency with two acousto-optic modulators in the beam preparation. One of the two beams (purple in Fig.~\ref{SLMPot}) is converted to a TEM10 mode using a spatial light modulator. To reduce drifts, the pointing and displacement of the TEM00 beam used for detection (red beam) are stabilized. Following the trap, a split-self homodyne detection reads out the motion along the inverted potential, while a single photodetector with a pinhole collects axial position information.

\section{Parameters required for the LQG implementation}

A key element of the optimal feedback scheme (LQG algorithm) is the accurate mathematical description of the experimental conditions. We divide the latter into four categories: the system dynamics, the read-out, the forcing apparatus, and the external noise processes.

For modelling the system dynamics (assumed linear), the LQG needs the experimental calibration of the \textbf{dissipation coefficient} $\gamma$ and of the \textbf{inverted stiffness} $k_{\mathrm{x,IP}}$. The first one is inferred from the thermal spectrum fit of the particle position in the harmonic tweezer (see Fig.~\ref{Spectrum_Detection}), while the $k_{\mathrm{x,IP}}$ is calibrated by fitting the bi-modal probability distribution of the free evolution in the inverted potential (like the one plotted in orange in Fig. 1a).
\SD{The actual value of the mass does not enter in the model, is fixed to a reasonable value from the particle solution specification, and is used for unit considerations.}{The mass value is given by the radius nominal value and the particle density: $m=\SI{9.0\pm 0.8e-18}{kg}$. Note that the functionality of the feedback is not affected by this parameter as it enters the model of the dynamics as a common scaling parameter.}

Regarding the readout calibration, the output equation must be identified, i.e. the relation between the displacement of the particle and the measured voltage. In the linear approximation, this corresponds to a linear combination of the individual axial displacements weighted with the \textbf{detection sensitivity}, i.e., the calibration factor converting the measured voltage in the detection channels ($x$ and $z$ detections) into the corresponding displacements in meters for all visible particle motion axes (therefore including crosstalks). The next Section \ref{SuppOpticalDetection} details the optical detection calibration process. In addition, we also calibrate the \textbf{variance of the measurement noise}, assumed to be white.

In the same vein, the LQG needs the calibration of the forcing apparatus, i.e. the calibrated value of the Newton-to-volt conversion operated by the control electrodes. The \textbf{force calibration} procedure is addressed in the dedicated Section~\ref{SuppForcecalib}. 

Finally, the LQG requires the \textbf{external process noises} acting on all components of the Kalman filter state vector. At our operating pressure, the principal noise driving the particle motion is gas collision induced thermal white noise, whose variance of $2k_BT_0\gamma$ is inferred from the fit of the harmonic spectrum. In the \textbf{adaptive} approach, the apex position $x_{\mathrm{apex}}$ is included in the state vector. Contrary to mechanical processes, we have very little information on the apex drift process noise. We, therefore, approximate it as a constant parameter excited by a white noise source of a magnitude consistent with the drift slow timescale.
As a consequence, the position estimate has a longer time constant than the mechanical states estimates driven by thermal noise. In other words, it is modified less aggressively in each time step of the Kalman filter (with a sampling rate $T_e=\SI{32}{ns}$).
Due to the lack of information on the apex position, process noise, and the assumptions made, the $x_{\mathrm{apex}}$ estimate requires hundreds of thousands of time steps to converge to a reliable value: the adaptation time constant (for a $95\%$ convergence) is worth $\tau=\SI{0.6}{ms}$. However, as long as the actual drifts of $x_{\mathrm{apex}}$ happen slower than the estimation time constant, the Kalman filter is able to track the drift sufficiently accurately to maintain stability.

\section{Optical detection}
 \label{SuppOpticalDetection}
 
\begin{figure}[h!]
	\centering
	\includegraphics[width=10cm]{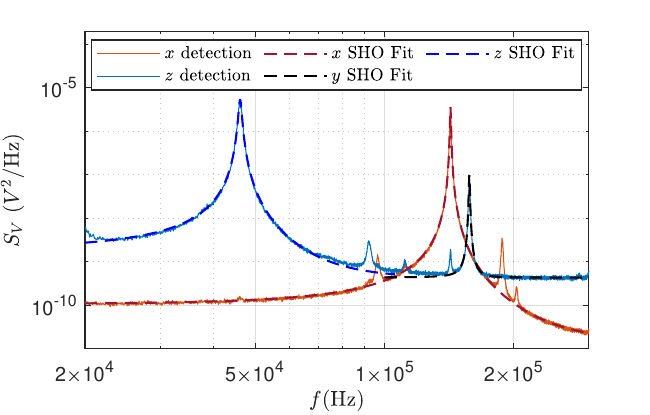}
	\caption{\textbf{Sup. Mat.: $x$ and $z$ detection signals spectrum in the harmonic potential (TEM10 beam switched off)}. Both are inferred from $\SI{3}{s}$ acquisition in the harmonic trap created by the TEM00 mode only (same $y$ and $z$ dynamics as in the inverted potential with the TEM10 switched on, but the $x$ motion is here harmonically resonant). All degrees of freedom are fitted by thermal harmonic oscillator spectrum (SHO) in dashed lines to calibrate the two detections sensitivities of the detection to each motional degree of freedom. }
	\label{Spectrum_Detection}
\end{figure}

The particle position is measured using a standard split-mirror detection scheme \cite{hebestreit_calibration_2018,Gieseler2012}. The forward propagating TEM00 light mode is separated from the TEM10 mode, attenuated, and split along the vertical axis by means of a sharp-edged D-shaped mirror for the $x$ detection and spatially filtered by a pinhole for the $z$ detection. Light is attenuated and sent to two pairs of balanced InGaAs detectors (PDB425C, Thorlabs Inc.) where the difference signals $\phi_x$ and $\phi_z$ are measured. The two forward detection channels displayed in Fig.~\ref{Spectrum_Detection} are called $x$ detection (orange) and $z$ detection (blue) because they show primarily the corresponding degree of freedom. By analysing the detection spectrum in the harmonic case shown in Fig.~\ref{Spectrum_Detection}, we can actually see that the $z$ detection shows a large $y$ cross-talk and a slight $x$ one, while the $x$ detection only includes a barely visible $z$ contribution. In other words, we can write: $\phi_z= C_{zz} x+ C_{zy}y$, with $C_{zy}< C_{zz}$ and $\phi_x= C_{xx} x$.

The aforementioned detection sensitivities and cross-talks are calibrated with the TEM10 mode set to zero. By assuming harmonic motion in the TEM00, we fit each degree of freedom by a thermal harmonic spectrum. We obtain $C_{xx}=\SI{2.7\pm 0.2 e6}{V/m}$, $C_{zz}=\SI{9.98\pm 0.02 e5}{V/m}$, $C_{zy}=\SI{4.5\pm 0.02e5}{V/m}$.

In a first approximation, the parameters are assumed constant, but we identify two main phenomena that affect them: 

\begin{enumerate}
\item The optical detection sensitivities are constant as long as the particle remains in the center of the TEM00 beam (stabilized by piezo controllers). Hence, if the stabilization starts failing and the particle exceeds the $\pm \SI{100}{nm}$ around the beam center, the calibrated values start being more than $20\%$ off, hence creating a model mismatch and damaging the efficiency of the feedback. We, therefore, ensure the particle is stabilized within this range.

\item When the orthogonally polarized TEM10 mode is turned on to create the inverted potential, the polarized beam-splitter placed before the detectors no longer filters it out perfectly from the TEM00. Due to the different intrinsic symmetries of the TEM10 beam, the $y$ and $z$ information encrypted in its scattering are emphasized by the split detection tuned for the TEM00 $x$ detection.
As a consequence, higher cross-talks with the $y$ and $z$ motion appear in the $x$ detection signal during stabilization on the inverted potential (as visible in Fig.~\ref{Spectrum}): $\phi_x^{\textrm{IP}}= C_{xx} x+ C_{xy}^{\textrm{IP}} y+C_{xz}^{\textrm{IP}}z$. With the calibration in the inverted configuration being much less reliable, we only estimate $C_{xy}^{IP}$ and $C_{xz}^{IP}$ to be worth $\sim\SI{7e5}{V/m}$. 
 
\end{enumerate}

Let us finally recall that while only the $\phi_x$ is used for the \textbf{non-adaptive} approach, the \textbf{adaptive} approach uses both $\phi_x$ and $\phi_z$ to estimate the $x$ (and $z$) motions. Including the $\phi_z$ independent detection channel indeed mitigates the impact of the above-mentioned imperfection of the $z$ cross-talk calibration ($C_{xz}^{IP}$) on the LQG efficiency. On the other hand, the error on the $C_{xy}^{IP}$ calibration does not deteriorate the LQG efficiency thanks to the frequency split between the $y$ resonance (higher frequency) and both the $x$ motion (low frequency stabilized motion), and the $z$ motion (lower resonance frequency $f_z$) frequency ranges.
.

\section{Force calibration}
\label{SuppForcecalib}
\SD{}{The electrostatic force is applied to the particle via a pair of electrodes of dimension 40x11x0.1mm in height width and length, separated by approximately 230$\mu m$.}
The force felt by the particle $\vec{f_e}$ is proportional to the voltage $u$ applied to the electrodes installed along the $x$ axis. Due to a slight misalignment, we measure a cross-coupling in the $z$ direction: $\vec{f_e}=f_{e,x} \vec{x}+ f_{e,z} \vec{z}=u\times (C_{u,x} \vec{x}+C_{u,z} \vec{z})$, with the $C_{u,.}$ being the force calibration factors.
The parameters are calibrated by applying a sinusoidal voltage $u=sin(2\pi f t)$ at different known frequencies $f$ and measuring the corresponding responses in the already calibrated detections\cite{magrini_supp_2021,zhu_nanoscale_2023,PhysRevA.106.043511}. Those factors depend on the charge of the particle and are worth $C_{u,x}=\SI{4.98\pm 0.02 e-13}{N/V}$ and $C_{u,z}=\SI{1.4\pm 0.1e-13}{N/V}$ for the results shown in this article. 

\section{Position power density spectrum of the stabilization on the inverted potential}

\begin{figure}[h!]
	\centering
	\includegraphics[width=10cm]{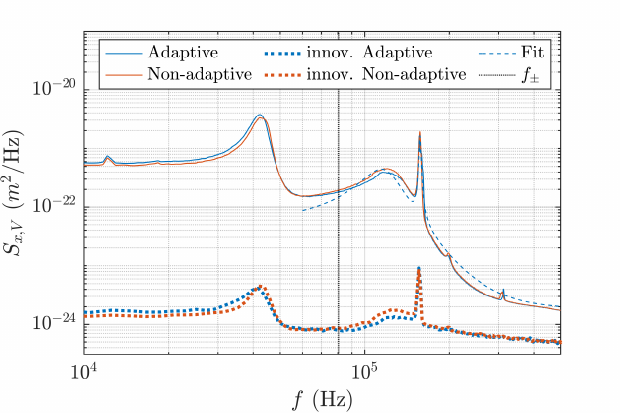}
	\caption{\textbf{Supp. Mat.: $x$ detection and innovation spectra during feedback stabilization.} All spectra are inferred from $\SI{5}{s}$ acquisition of the $x$ detection signal while the particle is stabilized on the apex of the inverted potential. Both non-adaptive and adaptive approaches (orange and blue plain lines) show cross-talks with the $y$ motion and the broadly excited $z$ resonance frequency range. The corresponding innovation spectrum (orange and blue dotted line) shows the error between the Kalman filters $x$ estimate and the $x$ measurement. The innovation sequence for the adaptive approach demonstrates a better estimate around the $x$ stabilized motion frequency range (around $\omega_\textrm{eff}=2\pi\times \SI{118}{kHz}$).}
	\label{Spectrum}
\end{figure}

Fig.~\ref{Spectrum} displays the Power Spectrum Density of the $x$ detection signal and the innovation signal (error between the Kalman estimate and the measurement) during the stabilization traces shown in Fig.~2.

The $x$ detection spectrum during both \textbf{non-adaptive} approach and \textbf{adaptive} approach stabilization shows some cross-talk with the other motional degrees of freedom. As the particle dynamics in $y$ and $z$ axes remain unchanged after switching between the inverted potential and the harmonic potential, we use the calibration of Fig.~\ref{Spectrum_Detection} and Table~1 to identify the origin of the visible resonant peaks.  The peak at $\SI{160}{kHz}$ corresponds to thermal $y$ motion in the TEM00 harmonic tweezer, only more visible due to the above-mentioned unwanted detection of the TEM10 scattering. Meanwhile, the broad peak at lower frequency is identified as the $z$ thermal motion, further excited (and deformed) by the low frequency driving applied to the $x$ electrodes with some cross-coupling to the $z$ direction. This is why using the second detection for estimating the particle position strongly benefits apex tracking and stabilization performance. 
	
To complement the temporal analysis carried out in the main article, we fit the stabilization spectral response by a double thermal spectrum, from which we extract the $x$ motion and ignore the frequencies associated with $z$ and $y$ cross-talks. We obtain $\omega_\textrm{eff}=2\pi\times \SI{118\pm 5}{kHz}$, $Q_{\textrm{eff}}=m\omega_\textrm{eff}/\gamma_{\textrm{eff}}=3\pm 0.5$, $T_{\textrm{eff}}=\SI{10\pm2}{K}$ and $\sigma_{\textrm{eff}}=\sqrt{\frac{k_B T_{\textrm{eff}}}{m\omega_\textrm{eff}^2}}=\SI{7\pm1}{nm}$. The spectrum fit values (with large error bars due to the modest quality of the fit) are consistent with the equipartition $T_{\textrm{eff}}=m\sigma_v^2/k_B=\SI{16\pm1}{K}$ and the position standard deviation $\sigma_x=\SI{9\pm0.5}{nm}$ inferred from the temporal study detailed in the article. The slight overestimation of the above is because neither $z$ nor $y$ motions are filtered out from the $x$ detection signal in the temporal analysis, whereas the spectral one allows the $x$ motion to be isolated in $\phi_x$. 

Note that the spectra of both the \textbf{non-adaptive} and the \textbf{adaptive} approach do not show a peak around the double-well harmonic resonance frequency ($f_{x,\pm}$ in dotted line), therefore meeting criterion 2.

\section{Stabilisation success indicators}
\label{choice_of_stab}
All three criteria complement each other to ensure the stability of the trajectory at the intensity minimum. In particular, those indicators allow us to exclude two artifacts that are misleading in the stabilization analysis:

\begin{enumerate}
    \item \textbf{Stabilization on the inverted potential slopes:} The particle may be stabilized on one of the slopes of the inverted potential rather than on the apex of the potential. This can occur, for example, if there are drifts in the alignment of the TEM10 beam with respect to the TEM00 beam, leading to a displacement of the inverted potential maximum $x_{\mathrm{apex}}$ away from the $x=0$ position. In such a scenario, criteria 1 and 3 may still be met, making the trajectory appear stable around $x=0$. However, it would require a constant non-zero-mean force to compensate for the potential slope, resulting in a non-zero average control driving (as in Fig.~2b)). Hence the importance of criterion 3.
    
    \item \textbf{Harmonic evolution in a local minimum of the inverted potential:} For example, the center of the inverted potential $x_{\mathrm{apex}}$ may be significantly out-centered, causing the position $x=0$ to coincide with the center of one of the double wells. The trajectory would hence appear stable. This scenario is less likely than the first one, and can be easily excluded by verifying criterion 3 in Fig.~\ref{Spectrum}. A particle harmonically trapped in one of the wells would exhibit the corresponding harmonic resonance frequency.
\end{enumerate}

\section{Adaptive apex tracking}

\begin{figure}[h!]
	\centering
	\includegraphics[width=11cm]{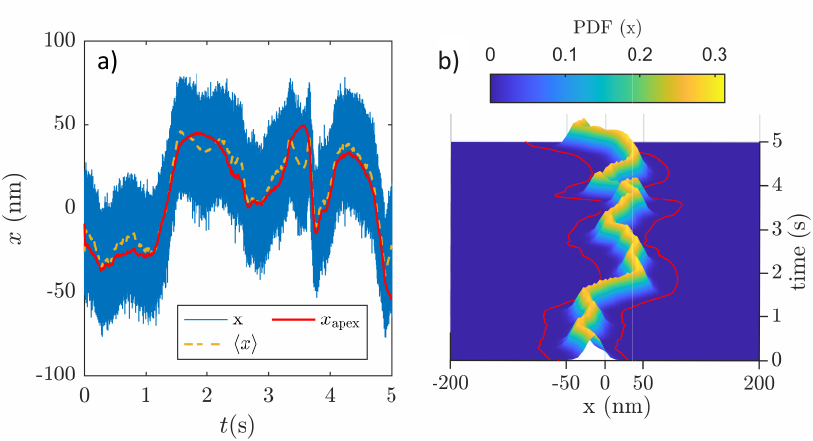}
	\caption{\textbf{Supp. Mat.: Example of apex position tracking by the feedback:} The \textbf{adaptive} approach uses the two independent detection channels to estimate the inverted potential apex position $x_{\mathrm{apex}}$ in real-time (with $\tau=\SI{0.6}{ms}$ convergence time). The driving is adjusted to stabilize the particle on the apex, following the small drifts happening on the $\SI{50}{ms}$ time scale in the potential shape due to beam positioning. (a) The $x$ detection in blue corresponds to the same trajectory as the one used for Fig.~2c, on which is superimposed the moving average in dashed yellow and the apex position estimate $x_{\mathrm{apex}}$ from the \textbf{adaptive} approach in red. (b) Alternative view of Fig.~2c in the position frame of reference. The red lines correspond to $x_{\mathrm{apex}}\pm \SI{50}{nm}$ from the \textbf{adaptive} approach $x_{\mathrm{apex}}$ real-time estimate.}
	\label{apex tracking}
\end{figure}

The \textbf{adaptive} approach tracks the position of the apex $x_{\mathrm{apex}}$ in real-time and adapts the control targeted stabilization position accordingly. The apex estimated position of the trajectory used for Fig.~2c is plotted in red in Fig.~\ref{apex tracking}a. Fig.~\ref{apex tracking}b is an alternative display of the position PDF shown in Fig.~2c, but in the detection frame of reference. This view emphasizes the drifts of the apex of the potential on which the particle is successfully stabilized. The 0-mean force driving shown in Fig.~2d indeed demonstrates the validity of apex tracking - the particle is stabilized at the meta-stable point (moving average of the position in yellow in Fig.~\ref{apex tracking}a following $x_{\mathrm{apex}}$, and PDF(x) confined within $x_{\mathrm{apex}}\pm \SI{50}{nm}$ red lines in Fig.~\ref{apex tracking}b).

\bibliography{Inverted_Supp}